\documentclass[pra, twocolumn, nofootinbib, eqsecnum, showpacs, showkeys]{revtex4}

\usepackage{graphics,amssymb}

\newcommand {\psig}{\Sigma_}
\newcommand {\pxi}{\Xi_}
\newcommand {\apsig}[1]{\langle \Sigma_{#1} \rangle}

\newcommand {\avecsig}{\langle{\mbox{\boldmath$\Sigma$}} \rangle}
\newcommand{\bsigma}{{\mbox{\boldmath$\Sigma$}}}

\newcommand{\ehat}{\hat{{\bf e}}}
\newcommand{\fhat}{\hat{{\bf f}}}
\newcommand{\nhat}{\hat{{\bf n}}}

\begin{document}

\title{Maps for Lorentz transformations of spin}

\author{Thomas F. Jordan}
\email[email: ]{tjordan@d.umn.edu}
\affiliation{Physics Department, University of Minnesota, Duluth, Minnesota 55812}
\author{Anil Shaji}
\email[email: ]{shaji@unm.edu}
\altaffiliation[Present address: ]{The University of New Mexico, Department of Physics and Astronomy, 800 Yale Blvd. NE, Albuquerque, New Mexico 87131}
\affiliation{The University of Texas at Austin, Center for Statistical Mechanics, 1 University Station C1609, Austin Texas 78712}
\author{E. C. G. Sudarshan}
\email[email: ]{sudarshan@physics.utexas.edu}
\affiliation{The University of Texas at Austin, Center for Statistical Mechanics, 1 University Station C1609, Austin Texas 78712}  

\begin{abstract}
Lorentz transformations of spin density matrices for a particle with positive mass and spin $1/2$ are described by maps of the kind used in open quantum dynamics. They show how the Lorentz transformations of the spin depend on the momentum. Since the spin and momentum generally are entangled, the maps generally are not completely positive and act in limited domains. States with two momentum values are considered, so the maps are for the spin qubit entangled with the qubit made from the two momentum values, and results from the open quantum dynamics of two coupled qubits can be applied. Inverse maps are used to show that every Lorentz transformation completely removes the spin polarization, and so completely removes the information, from a number of spin density matrices. The size of the spin polarization that is removed is calculated for particular cases.    
\end{abstract}

\date{\today}


\maketitle

\section{Introduction}\label{intro}
Quantum information is not independent of relativity. The spin of a particle with mass, for example an electron spin, is changed by Lorentz transformations. If the spin is used to handle a qubit of quantum information, that information will be changed by Lorentz transformations \cite{czachor97a,peres02a,alsing02a,gingrich02a,ahn03a,alsing03a,peres03a,gonera04a,bartlett05a}. This may not be important for the effort to realize the promise of quantum computing \cite{nielsen00,preskill04b}. Particles moving at relativistic speeds are not likely to be needed for quantum information processing devices. Nevertheless, questions of relativity may be interesting for those who look for deeper understanding of the nature of quantum information. Understanding can grow with experience as work ranging from theoretical foundations to practical applications involves various properties of quantum information. This work on relativistic properties is meant to be one contribution. Related developments are reviewed by Peres and Terno \cite{peres04a}.

We consider Lorentz transformation of spin as an example of the kind of map used in open quantum dynamics. There the map describes evolution of density matrices for a subsystem caused by unitary Hamiltonian evolution in a larger system \cite{sudarshan61a,jordan61,jordan62,davies76,kraus83,alicki87,breuer02,zyczkowski04a}. Here we consider the map that describes the transformation of the spin density matrix caused by a unitary Lorentz transformation of the state of a particle with spin and momentum. The spin plays the role of a subsystem in the larger system described by spin and momentum. With this map we can move beyond the statement that there is no Lorentz transformation of spin that is independent of momentum \cite{peres02a} and say exactly how the Lorentz transformation of the spin depends on the momentum.  We can say that several different ways by using different forms of the map. 

We consider spin $1/2$, consider states where the momentum is concentrated around two different values, and make a qubit with the two momentum values. Then we have two qubits, the spin and the one made from the momentum, and the map is like those we have described for the dynamics of two coupled qubits \cite{jordan04}. The spin and momentum generally are entangled, so the map generally is not completely positive and acts on a limited domain \cite{pechukas94,stelmachovic01a,jordan04}. This domain is exactly the same as the one we described in detail for the first example of dynamics we considered \cite{jordan04}. It tells us the bounds on the variables in our equations. In other respects, the map that describes Lorentz transformation of spin is very different from those we have considered for dynamics and provides a new and helpful illustration of the way these maps work \cite{jordan05b}. 

Our results complement those that have been presented for a momentum distribution that is maximum at one value \cite{peres02a,gonera04a}. We use inverse maps to show that every Lorentz transformation completely removes the spin polarization, and so completely removes the information, from a number of spin density matrices. We calculate the size of the spin polarization that is removed in particular cases.

Except for questions of domains, generalization to any finite number of momentum values is easy. Results can be inferred.  
\vspace{0 cm}

\section{Lorentz transformations of spin} \label{sec1}

We consider a particle with positive mass and spin $1/2$. We use Pauli matrices $\psig 1$, $\psig 2$, $\psig 3$ to represent the spin as $\frac{1}{2} \bsigma$. Let $|p,\, s \rangle$ be the eigenvectors for the eigenvalues $p=(p_0,\, {\bf p})$ of the four-momentum and $s= \pm \frac{1}{2}$ for the 3 component of the spin. The unitary operator $U(\Lambda)$ for a Lorentz transformation $\Lambda$ gives \cite{weinberg95}
\begin{equation}
  \label{eq:lt1}
  U(\Lambda) |p \,,\, s \rangle = \sqrt{\frac{(\Lambda p)^0}{p^0}} \sum_{s'=-1/2}^{1/2} D_{s's}(W(\Lambda, p))|\Lambda p \,,\, s' \rangle
\end{equation}
where $W(\Lambda,\, p)$ is a rotation, the Wigner rotation, which depends on $\Lambda$ and $p$, and $D(W)$ for a rotation $W$ is the $2$$\times$$2$ unitary rotation matrix made from the $\psig j$ so that 
\begin{equation}
  \label{eq:lt1a}
  D(W)^{\dagger} \bsigma D(W) = W(\bsigma) 
\end{equation}
where $W(\bsigma)$ is simply the vector $\bsigma$ rotated by $W$.  

A pure state for the particle is represented by a vector
\begin{equation}
  \label{eq:stateini}
  |\Psi \rangle = \sum_{s=-1/2}^{1/2} \int d^3 {\bf p}\; \Psi_{s}({\bf p})| p,\, s \rangle.
\end{equation}
The reduced density matrix for the spin for this state is
\begin{widetext}
\begin{eqnarray}
  \label{eq:spinini}
  \rho & = & {\mbox{Tr}}_{\bf p} [| \Psi \rangle \langle \Psi | \,] = \int d^3 {\bf p} \langle {\bf p} | \Psi \rangle \langle \Psi | {\bf p} \rangle \nonumber \\
       & = & \sum_{s_1,s_2}\int d^3 {\bf p} \; \int d^3 {\bf p}_1 \; \int d^3 {\bf p}_2 \; \delta^{(3)}({\bf p}-{\bf p}_1) \delta^{(3)}({\bf p}-{\bf p}_2) \; \Psi_{s_1}({\bf p}_1) |s_1\rangle \langle s_2|\Psi^*_{s_2}({\bf p}_2) \nonumber \\
       & = &  \int d^3 {\bf p} \widetilde{\rho} ({\bf p})
\end{eqnarray}
where
\begin{equation}
\label{eq:rhotilde}
 \widetilde{\rho}_{s_1s_2}({\bf p}) \equiv \Psi_{s_1}({\bf p}) \Psi^*_{s_2}({\bf p}).
\end{equation}
A Lorentz transformation $\Lambda$ changes the state vector $|\Psi \rangle$ to
\begin{equation}
  \label{eq:finstate}
  |\Psi^{\Lambda}\rangle =U(\Lambda)|\Psi \rangle= \sum_s \int d^3 {\bf p} \; \sqrt{\frac{(\Lambda p)^0}{p^0}} \Psi_{s}({\bf p}) \sum_{s'} D_{s's}(W(\Lambda, p))|\Lambda p, \, s' \rangle.
\end{equation}
and the spin density matrix to
\begin{eqnarray}
  \label{eq:rhofin}
  \rho^{\Lambda} &=& {\mbox{Tr}}_{\bf p} [| \Psi^{\Lambda} \rangle \langle \Psi^{\Lambda} | \,] \nonumber \\
  & = & \sum_{s_1,s_2,s_3,s_4} \int d^3 {\bf p} \; \int d^3 {\bf p}_1 \; \int d^3 {\bf p}_2 \; \sqrt{\frac{(\Lambda p_1)^0}{(p_1)^0} \frac{(\Lambda p_2)^0}{(p_2)^0}} \delta^{(3)}({\bf p} - \Lambda {\bf p}_1)\delta^{(3)}({\bf p} - \Lambda {\bf p}_2) \nonumber \\
 && \hspace{3 cm} \times \quad D_{s_3s_1}(W(\Lambda, \, p_1)) \Psi_{s_1}({\bf p}_1) |s_3 \rangle \langle s_4| \Psi^*_{s_2}({\bf p}_2) [D_{s_4s_2}(W(\Lambda, \, p_2))]^{*} \nonumber \\
 & = & \int d^3 {\bf p} \;  D(W(\Lambda, \, p)) \; \widetilde{\rho}({\bf p}) \; [D(W(\Lambda, \, p))]^{\dagger} .
\end{eqnarray}
Here we used
\begin{equation}
 \delta^{(3)}(\Lambda {\bf p}_1 - \Lambda {\bf p}_2) = \frac{p_1^0}{(\Lambda p_1)^0} \delta^{(3)}({\bf p}_1-{\bf p}_2). 
\end{equation}
The results expressed in terms of density matrices hold for mixed states of the particle as well as for pure states; linear combinations of the equations for different pure-state vectors $| \Psi \rangle$ give the same equations for mixed states.  

\vspace{5 cm}

\end{widetext}

\section{Two momentum values} \label{sec2}

We consider just two different momentum values ${\bf p}_1$ and ${\bf p}_2$. Let
\begin{equation}
  \label{eq:two1}
  \rho = \widetilde{\rho}_1 + \widetilde{\rho}_2
\end{equation}
with $\widetilde{\rho}_1$ for ${\bf p}$ equal to (or concentrated around) ${\bf p}_1$ and $\widetilde{\rho}_2$ for ${\bf p}$ equal to (or concentrated around) ${\bf p}_2$. Let
\begin{eqnarray}
  \label{eq:two2}
  \widetilde{\rho}_1 & = & \frac{1}{2} \left( q + {\bf r}_1 \cdot \bsigma \right) \nonumber \\ \nonumber \\
  \widetilde{\rho}_2 & = & \frac{1}{2} \left(1- q + {\bf r}_2 \cdot \bsigma \right)
\end{eqnarray}
with $q$ between $0$ and $1$. Then
\begin{equation}
  \label{eq:two3}
  \langle \bsigma \rangle = {\mbox{Tr}} \left[ \bsigma \rho \right] = {\bf r}_1 + {\bf r}_2.
\end{equation}
We assume the momentum is concentrated closely enough that we can use ${\bf p}_1$ for ${\bf p}$ in the Wigner rotations for $\widetilde{\rho}_1$ and use ${\bf p}_2$ for ${\bf p}$ in the Wigner rotations for $\widetilde{\rho}_2$ and accept the accuracy of that approximation. Then
\begin{equation}
  \label{eq:two4}
  \rho^{\Lambda} = D(W_1) \widetilde{\rho}_1 D(W_1)^{\dagger} +  D(W_2) \widetilde{\rho}_2 D(W_2)^{\dagger}
\end{equation}
or
\begin{equation}
  \label{eq:two5}
  \langle \bsigma \rangle^{\Lambda} = {\mbox{Tr}} \left[ \bsigma \rho^{\Lambda} \right] = W_1 ({\bf r}_1) + W_2 ({\bf r}_2)
\end{equation}
where
\begin{equation}
  \label{eq:two6}
  W_1 = W(\Lambda, \, p_1) \quad , \quad W_2 = W(\Lambda, \, p_2)
\end{equation}
and each $W({\bf r})$ is simply the vector ${\bf r}$ rotated by $W$. Generalization of these equations to any finite number of momentum values is obvious.

Let $\pxi 1$ be a matrix made from the momentum operator that has eigenvalue $1$ when ${\bf p}$ is (in the range of concentration around) ${\bf p}_1$ and eigenvalue $-1$ when ${\bf p}$ is (in the range of concentration around) ${\bf p}_2$. Then (in the approximation we are making)
\begin{equation}
  \label{eq:two7}
  {\bf r}_1 = \left \langle \bsigma \frac{1}{2} ( 1 + \pxi 1 ) \right \rangle \quad , \quad {\bf r}_2 = \left \langle \bsigma \frac{1}{2} ( 1 - \pxi 1 ) \right \rangle
\end{equation}
and
\begin{eqnarray}
  \label{eq:two8}
  \langle \bsigma \rangle^{\Lambda} & = & \frac{1}{2} \left[ W_1 (\langle \bsigma \rangle ) + W_2 (\langle \bsigma \rangle ) \right] \nonumber \\
 && \hspace{3 mm}+ \frac{1}{2} \left[ W_1 (\langle \bsigma \pxi 1 \rangle) - W_2 ( \langle \bsigma \pxi 1 \rangle ) \right].
\end{eqnarray}
The last term is 
\begin{equation}
  \label{eq:two9}
  \frac{1}{2} W_1 \left( \langle \bsigma \pxi 1 \rangle - W_1^{-1} W_2 \left( \langle \bsigma \pxi 1 \rangle \right) \right).
\end{equation}
The component of $\langle \bsigma \pxi 1 \rangle$ along the axis of $W_1^{-1}W_2$ drops out, so $\langle \bsigma \rangle^{\Lambda}$ depends only on the components of $\langle \bsigma \pxi 1 \rangle$ perpendicular to the axis of $W_1^{-1}W_2$. 

A spin state is described by the mean values $\langle \bsigma \rangle$, which are expressed in terms of ${\bf r}_1$ and ${\bf r}_2$ by Eq.(\ref{eq:two3}). A Lorentz transformation produces a map of spin states described by $\langle \bsigma \rangle$ changing to $\langle \bsigma \rangle^{\Lambda}$, given by Eq.(\ref{eq:two5}) or (\ref{eq:two8}). From Eq.(\ref{eq:two8}) we see that the map is described by $W_1$, $W_2$ and the components of
\begin{equation}
  \label{eq:two9ab}
  \langle \bsigma \Xi_1 \rangle = {\bf r}_1 - {\bf r}_2
\end{equation}
perpendicular to the axis of $W_1^{-1}W_2$. Both the spin state and the map are described completely by ${\bf r}_1$, ${\bf r}_2$, $W_1$, $W_2$. The Lorentz transformation and the two momentum values ${\bf p}_1$, ${\bf p}_2$ determine $W_1$ and $W_2$. The state of the particle determines $\langle \bsigma \rangle$ and $\langle \bsigma \Xi_1 \rangle$, or ${\bf r}_1$ and ${\bf r}_2$.

We assume that $\langle \bsigma \rangle$ and $\langle \bsigma \pxi 1 \rangle$ are mean values for a state of the particle. Our equations describe the Lorentz transformation of the spin only when $\langle \bsigma \rangle$ is compatible with the components of $\langle \bsigma \pxi 1 \rangle$ perpendicular to the axis of $W_1^{-1}W_2$ in describing a state of the particle. To see the restrictions that this implies, we treat the two momentum values as a qubit, so we have two qubits, the spin and the one made from the momentum. The linear map of matrices that describes the Lorentz transformation of density matrices $\rho$ for the spin is like the maps that we have used to describe the dynamics of two entangled qubits \cite{jordan04}. These maps generally are not completely positive and act in limited domains. Here the map that describes the Lorentz transformation is made to be used for the set of $\langle \bsigma \rangle$ that are compatible with the components of $\langle \bsigma \pxi 1 \rangle$ perpendicular to the axis of $W_1^{-1}W_2$. If the $3$ axis is taken to be along the axis of $W_1^{-1}W_2$, this set of compatible $\langle \bsigma \rangle$ is exactly the same as the compatibility domain that we described completely and precisely for an example of the dynamics of two entangled qubits \cite{jordan04}. It tells us the $\langle \bsigma \rangle$ and  $\langle \bsigma \pxi 1 \rangle$, or ${\bf r}_1$ and ${\bf r}_2$, for which our equations apply.

Explicitly, these are the ${\bf r}_1$ and ${\bf r}_2$ for which
\begin{equation}
  \label{eq:two9a}
  \sqrt{(r_1)^2 + (z_1)^2} + \sqrt{(r_2)^2 + (z_2)^2} \leq \sqrt{1 - (z_1+z_2)^2}
\end{equation}
where $r_1$ and $r_2$ are the lengths of ${\bf r}_1$ and ${\bf r}_2$ and $z_1$ and $z_2$ are the components of ${\bf r}_1$ and ${\bf r}_2$ along the axis of $W_1^{-1}W_2$, which we call the the $3$ axis. We find this limit by looking at
\begin{eqnarray}
  \label{eq:two9b}
  \avecsig & = & 2 {\bf r}_1 - \langle \bsigma \pxi 1 \rangle \nonumber \\ \nonumber \\
  \avecsig & = & 2 {\bf r}_2 + \langle \bsigma \pxi 1 \rangle
\end{eqnarray}
projected onto planes of constant $\apsig{3}$. For each fixed $\apsig{3}$, the compatibility domain is the set of $\avecsig$ in the elliptical area of the plane of fixed $\apsig{3}$ described \linebreak \cite[Eq.(2.77)]{jordan04} by, 
\begin{equation}
  \label{eq:two9c}
  d_1 + d_2 \leq 2 \sqrt{1 - \apsig{3}^2} 
\end{equation}
where $d_1$ and $d_2$ are the distances from the point $\avecsig$ to focii at the points in the plane described by the projections of $\pm \langle \bsigma \pxi 1 \rangle$. From Eqs.(\ref{eq:two9b}) we see that $d_1$ and $d_2$ are the lengths of the projections of $2 {\bf r}_1$ and $2 {\bf r}_2$, so the elliptical area is described by the inequality (\ref{eq:two9a}).

 We will stay within these bounds as we look at particular cases. We do not know what the similar restrictions would be for three or more momentum values. 

For the case of two momentum values considered here, we have shown \cite{jordan05b} that the linear map of matrices that describes the Lorentz transformation of density matrices $\rho$ for the spin is completely positive if and only if
\begin{equation}
  \label{eq:two10}
  W_1(\langle \bsigma \Xi_1 \rangle) = W_2 (\langle \bsigma \Xi_1 \rangle).
\end{equation}
Then $\langle \bsigma \Xi_1 \rangle$ is not changed by $W_1^{-1}W_2$, which means $\langle \bsigma \Xi_1 \rangle$ has no components perpendicular to the axis of $W_1^{-1}W_2$, so the compatibility domain is the set of all $\langle \bsigma \rangle$ for all spin states. Then also Eq.(\ref{eq:two8}) reduces to
\begin{equation}
  \label{eq:two11}
  \langle \bsigma \rangle^{\Lambda} = \frac{1}{2} W_1 \left( \langle \bsigma \rangle \right) + \frac{1}{2} W_2 \left( \langle \bsigma \rangle \right)
\end{equation}
so $\langle \bsigma \rangle^{\Lambda}$ is the same as it would be for a state of the particle represented by a density matrix that is a product of a density matrix for the spin and a density matrix for the momentum with equal probabilities for the two momentum values.

\section{Inverse maps} \label{inv}

Consider a state of the particle with two momentum values $p_1$ and $p_2$ and spin density matrix $\rho$ described by Eqs.(\ref{eq:two1}) and (\ref{eq:two2}). From this state, the Lorentz transformation $\Lambda$ produces a state with two momentum values $\Lambda p_1$ and $\Lambda p_2$ and spin density matrix
\begin{equation}
  \label{eq:inv1}
  \rho^{\Lambda} = \widetilde{\rho}_1^{\Lambda} + \widetilde{\rho}_2^{\Lambda}
\end{equation}
with
\begin{eqnarray}
  \label{eq:inv2}
  \widetilde{\rho}_1^{\Lambda} & = & \frac{1}{2} \left( q + W_1 ({\bf r}_1) \cdot \bsigma \right) \nonumber \\
  \widetilde{\rho}_2^{\Lambda} & = & \frac{1}{2} \left( 1-q + W_2 ({\bf r}_2) \cdot \bsigma \right).
\end{eqnarray}
We see that if ${\bf r}_1$ and ${\bf r}_2$ are for a state of the particle, then so are $W_1({\bf r}_1)$ and $W_2({\bf r}_2)$; they are for the Lorentz-transformed state, which the inverse Lorentz transformation takes to the state with ${\bf r}_1$ and ${\bf r}_2$. If the map from ${\bf r}_1$ and ${\bf r}_2$ to $W_1({\bf r}_1)$ and $W_2({\bf r}_2)$ describes a Lorentz transformation of the spin for a state of the particle, then so does the map from $W_1({\bf r}_1)$ and $W_2({\bf r}_2)$ to ${\bf r}_1$ and ${\bf r}_2$. We need to consider questions of domains only once, for ${\bf r}_1$ and ${\bf r}_2$, not again for $W_1({\bf r}_1)$ and $W_2({\bf r}_2)$.

Explicitly, the same inequality (\ref{eq:two9a}) that describes the compatibility domain for the map where $W_1$ and $W_2$ act on ${\bf r}_1$ and ${\bf r}_2$ also describes the compatibility domain for the inverse map where $W_1^{-1}$ and $W_2^{-1}$ act on $W_1 ({\bf r}_1)$ and $W_2 ({\bf r}_2)$. To see this, let $\hat{{\bf z}}$ be the unit vector along the axis of $W_1^{-1}W_2$, so that
\begin{equation}
  \label{eq:inv2a}
  W_1^{-1}W_2 (\hat{{\bf z}}) = \hat{{\bf z}}.
\end{equation}
Let $\hat{{\bf z}}'= W_1 (\hat{{\bf z}})$. Then
\begin{equation}
  \label{eq:inv2b}
  W_2 (\hat{{\bf z}}) = W_1 (\hat{{\bf z}}) = \hat{{\bf z}}',
\end{equation}
and
\begin{equation}
  \label{eq:inv2c}
  W_1 W_2^{-1} (\hat{{\bf z}}') = W_1 W_2^{-1}W_2 (\hat{{\bf z}}) = W_1 (\hat{{\bf z}}) = \hat{{\bf z}}',
\end{equation}
so $\hat{{\bf z}}'$ is a unit vector along the axis of $W_1W_2^{-1}$. In the inequality (\ref{eq:two9a}) the $r_1$ and $r_2$ are the lengths of $W_1 ({\bf r}_1)$ and $W_2 ({\bf r}_2)$ as well as the lengths of ${\bf r}_1$ and ${\bf r}_2$, and
\begin{eqnarray}
  \label{eq:inv2d}
  z_1 = \hat{{\bf z}} \cdot {\bf r}_1 & = W_1 (\hat{{\bf z}}) \cdot W_1 ({\bf r}_1) & = \hat{{\bf z}}' \cdot W_1 ({\bf r}_1) \nonumber \\ \nonumber \\
  z_2 = \hat{{\bf z}} \cdot {\bf r}_2 & = W_2 (\hat{{\bf z}}) \cdot W_2 ({\bf r}_2) & = \hat{{\bf z}}' \cdot W_2 ({\bf r}_2).
\end{eqnarray}

Suppose $\langle \bsigma \rangle=0$. From our description of the compatibility domain in the context of dynamics \cite{jordan04}, we know that zero $\langle \bsigma \rangle$ is compatible with any $\langle \bsigma \pxi 1 \rangle$ that is for a state of the particle. This requires only that $|\langle \bsigma \pxi 1 \rangle|$ be not larger than $1$. There is a state of the particle for any  ${\bf r}_1$ and ${\bf r}_2$ such that
\begin{equation}
  \label{eq:inv3}
  {\bf r}_1 + {\bf r}_2 = 0 \quad , \quad |{\bf r}_1 - {\bf r}_2| \leq 1.
\end{equation}
From $\langle \bsigma \rangle=0$, a Lorentz transformation can produce any
\begin{equation}
  \label{eq:inv4}
  \langle \bsigma \rangle^{\Lambda} = W_1 ({\bf r}_1) - W_2 ( {\bf r}_1)
\end{equation}
for $r_1 \leq 1/2$, and the inverse Lorentz transformation can change this $\langle \bsigma \rangle^{\Lambda}$ to zero. Since every Lorentz transformation is the inverse $\Lambda^{-1}$ of a Lorentz transformation $\Lambda$, we see that every Lorentz transformation completely removes the spin polarization $\avecsig^{\Lambda}$, and so completely removes the information, from a number of spin density matrices. We will see how big this $\langle \bsigma \rangle^{\Lambda}$ can be in particular cases in Section \ref{examples}.

\section{Wigner rotations}\label{wigner}

To work out examples we need to calculate Wigner rotations. Halpern \cite{halpern68} has calculated $D(W(\Lambda,p))$ for any $\Lambda$ and $p$. If
\begin{equation}
  \label{eq:rot3}
  p = (m \cosh \beta\, , \, m \sinh \beta \, \fhat).
\end{equation}
and $\Lambda$ is the Lorentz transformation for velocity \linebreak $v=\tanh \alpha$ in the direction of $\ehat$, then
\begin{equation}
  \label{eq:lo1}
  D(W(\Lambda,\, p))=\cos \frac{\varphi}{2} + i \sin \frac{\varphi}{2} (\bsigma \cdot \nhat). 
\end{equation}
where
\begin{eqnarray}
  \label{eq:rot2}
  \cos \frac{\varphi}{2} &\!\!\!\!=&\! \! \! \!  \frac{\cosh \frac{\alpha}{2} \cosh \frac{\beta}{2} + \sinh \frac{\alpha}{2} \sinh \frac{\beta}{2} \, (\ehat \cdot \fhat )}{ \sqrt{ \frac{1}{2} + \frac{1}{2} \cosh \alpha \cosh \beta + \frac{1}{2} \sinh \alpha \sinh \beta \, (\ehat \cdot \fhat)}}, \nonumber \\
  \sin \frac{\varphi}{2} \, \nhat & \!\!\!= &\! \! \! \frac{ \sinh \frac{\alpha}{2} \sinh \frac{\beta}{2} \, (\ehat \times \fhat ) }{\sqrt{\frac{1}{2} + \frac{1}{2} \cosh \alpha \cosh \beta + \frac{1}{2} \sinh \alpha \sinh \beta \, (\ehat \cdot \fhat)}}. \nonumber \\
\end{eqnarray}
This tells us both $D(W(\Lambda, \, p))$ and $W(\Lambda,\, p)$. The axis of the Wigner rotation $W(\Lambda, \, p)$ is $\nhat$ and the angle of $W(\Lambda,\, p)$ is $\varphi$.

\section{Examples} \label{examples}

To see large effects with simple examples, we let ${\bf p}_1$ and ${\bf p}_2$ be along the same line, which we call the $1$ axis, perpendicular to the direction of the Lorentz transformation, which we call the $2$ axis. Then both Wigner rotations $W_1$ and $W_2$ are around the $3$ axis and the angles $\varphi$ of the Wigner rotations are as big as they can be for given momentum magnitudes and Lorentz-transformation velocity. We let ${\bf r}_1$ and ${\bf r}_2$ be in the $1$, $2$ plane so they are fully exposed to the Wigner rotations.

In these cases, the linear map of density matrices that describes the Lorentz transformation of density matrices $\rho$ for the spin is completely positive if and only if either ${\bf p}_1 = {\bf p}_2$ or ${\bf r}_1 = {\bf r}_2$, because if ${\bf p}_1$ and ${\bf p}_2$ are different, then $W_1$ and $W_2$ are different and Eq.(\ref{eq:two10}) can be satisfied only if $\langle \bsigma \Xi_1 \rangle$, which is ${\bf r}_1 - {\bf r}_2$, is zero. 

\begin{widetext}

\begin{figure}[!hb]
   \begin{center}
     \includegraphics{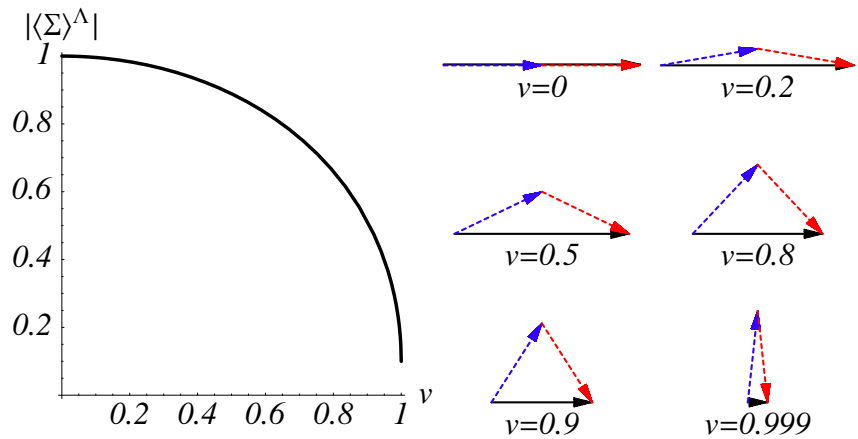}
   \end{center} \vspace{-10 mm}
   \caption{(color online)$\left| \langle \bsigma \rangle^{\Lambda} \right|$ as a function of the velocity $v$ of the Lorentz transformation, for ${\bf p}_1=-{\bf p}_2$ perpendicular to the direction of the Lorentz transformation and $|{\bf p}_1|/m=10$, with ${\bf r}_1$ and ${\bf r}_2$ perpendicular to the axes of the Wigner rotations, $r_1$ and $r_2$ both $1/2$, and ${\bf r}_1$ in the same direction as ${\bf r}_2$. On the right are the vectors $W_1({\bf r}_1)$ (blue dotted arrows), $W_2({\bf r}_2)$ (red dashed arrows), and $\langle \bsigma \rangle^{\Lambda}$ (black solid arrows) for different values of $v$.}
\label{fig1}
\end{figure} \vspace{-10 mm}
\begin{figure}[!htb]
   \begin{center}
     \includegraphics{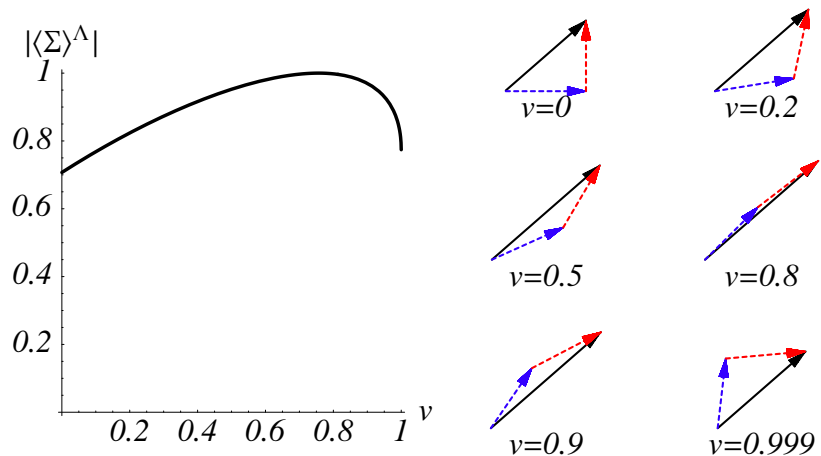}
   \end{center} \vspace{-10 mm}
   \caption{(color online) The same as in Fig. \ref{fig1} except that ${\bf r}_1$ is perpendicular to ${\bf r}_2$}
\label{fig2}
\end{figure}

\vspace{3 cm}

\end{widetext}

Let $\varphi_1$ and $\varphi_2$ be the angles of $W_1$ and $W_2$. Then the angle of $W_1^{-1}W_2$ is $\varphi_2 - \varphi_1$. If the angle between ${\bf r}_1$ and ${\bf r}_2$ is $\chi$, the angle between ${\bf r}_1$ and $W_1^{-1}W_2({\bf r}_2)$ is $\chi + \varphi_2 - \varphi_1$ and, from Eq.(\ref{eq:two5}), 
\begin{eqnarray}
  \label{eq:ex1}
  \left| \langle \bsigma \rangle^{\Lambda} \right| & = & \left| {\bf r}_1 + W_1^{-1}W_2({\bf r}_2 ) \right| \nonumber \\
 & = & \sqrt{r_1^2 + r_2^2 + 2 r_1 r_2 \cos ( \chi + \varphi_2 - \varphi_1) }. \nonumber \\
\end{eqnarray}
This is maximum when $\chi+ \varphi_2 - \varphi_1$ is zero. If the linear map of matrices that describes the Lorentz transformation of density matrices $\rho$ for the spin is completely positive, then ${\bf r}_1={\bf r}_2$ and $\chi$ is zero. Then the Lorentz transformation can only decrease $\left| \langle \bsigma \rangle \right|$ as it makes $\varphi_2$ and $\varphi_1$ nonzero. This decreasing $\left| \langle \bsigma \rangle^{\Lambda} \right|$ is shown in \linebreak Fig. \ref{fig1} as a function of the velocity $v$ of the Lorentz transformation for the case where $r_1$ and $r_2$ are $1/2$ and $\chi$ is zero, $|{\bf p}_1|/m$ is 10 and ${\bf p}_2$ is $-{\bf p}_1$. When $\chi$ is not zero, $|\langle \bsigma \rangle|$ can be increased as well as decreased by the Lorentz transformation. This is shown in Fig. \ref{fig2} where everything is the same as in Fig. \ref{fig1} except that $\chi$ is $- \pi/2$.

To see the size of the $\left| \langle \bsigma \rangle \right|$ that a Lorentz transformation can completely remove, or produce starting from zero, we reproduce Eq.(\ref{eq:inv4}) by letting ${\bf r}_2$ be $-{\bf r}_1$, which means letting $r_2$ and $\chi$ be $r_1$ and $\pi$ in Eq.(\ref{eq:ex1}). The $\left| \langle \bsigma \rangle^{\Lambda} \right|$ described by Eq.(\ref{eq:inv4}) is shown in Fig. \ref{fig3} as a function of the velocity $v$ of the Lorentz transformation for different values of ${\bf p}_1 \cdot {\bf p}_2/ |{ \bf p}_1 |^2$, with ${\bf p}_1$ and ${\bf p}_2$ still along the same line, for the case where $r_1$ is $1/2$ and $|{\bf p}_1|/m$ is $10$ as before.

\begin{widetext}

\vspace{1 cm}
  
\begin{figure}[!hb]
   \begin{center}
     \includegraphics{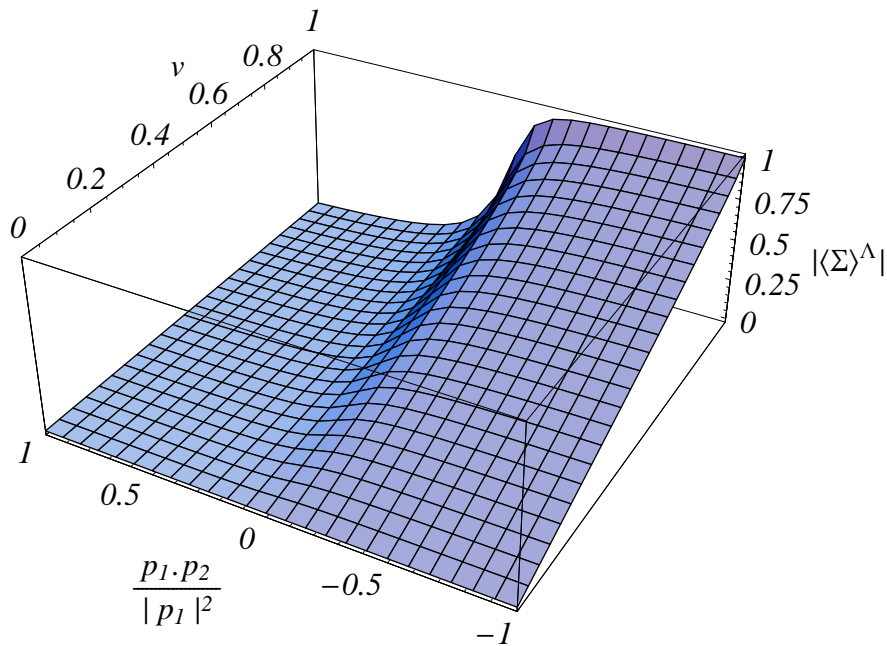}
   \end{center} \vspace{1 cm}
   \caption{(color online) The size of the $\langle \bsigma \rangle^{\Lambda}$ described by Eq.(\ref{eq:inv4}), that a Lorentz transformation can completely remove, or produce starting from zero, as a function of the velocity $v$ of the Lorentz transformation, for different values of ${\bf p}_1 \cdot {\bf p}_2/ | {\bf p}_1 |^2$ with ${\bf p}_1$ and ${\bf p}_2$ along the same line perpendicular to the direction of the Lorentz transformation and ${\bf p}_1/m=10$, with ${\bf r}_1$ perpendicular to the axis of the Wigner rotation and $r_1 = 1/2$. }
\label{fig3}
\end{figure}

\end{widetext}

\bibliography{ncp}

\end{document}